\documentclass[12pt]{article}

\usepackage{amsmath}
\usepackage{amssymb}
\usepackage{graphicx}%
\usepackage{authblk}

\begin{document}

\title{Scalar Field Cosmology II: Superfluidity, Quantum Turbulence, and Inflation}

\author[ab]{Kerson Huang\thanks{kerson@mit.edu}}

\author[b]{Hwee-Boon Low}

\author[b]{Roh-Suan Tung\thanks{rohsuan.tung@ntu.edu.sg}}

\affil[a]{\it Physics Department, Massachusetts Institute of Technology,
Cambridge, MA, USA 02139}

\affil[b]{\it Institute of Advanced Studies,
Nanyang Technological University, Singapore 639673}

\date{}

\maketitle

\begin{abstract}
We generalize the big-bang model in a previous paper by extending the real
vacuum scalar field to a complex vacuum scalar field, within the FLRW
framework. The phase dynamics of the scalar field, which makes the universe a
superfluid, is described in terms of a density of quantized vortex lines,
and a tangle of vortex lines gives rise to quantum turbulence. We propose that
all the matter in the universe was created in the turbulence, through
reconnection of vortex lines, a process necessary for the maintenance of the
vortex tangle. The vortex tangle grows and decays, and its lifetime is the era
of inflation. These ideas are implemented in a set of closed cosmological
equations that describe the cosmic expansion driven by the scalar field on
the one hand, and the vortex-matter dynamics on the other. We show how these
two aspects decouple from each other, due to a vast difference in energy
scales. The model is not valid beyond the inflation era, but the universe
remains a superfluid afterwards. This gives rise to observable effects in the
present universe, including dark matter, galactic voids, non-thermal
filaments, and cosmic jets.
\end{abstract}

%{PACS numbers: 98.80.Bp, 98.80.-k, 03.70.+k, 98.80}

\section{Introduction and summary}

In paper I of this series [1], we describe the emergence of a vacuum scalar
field at the big bang, using a Halpern-Huang quantum scalar field [2] in the
FLRW (Friedman-Lamaitre-Robertson-Walker) framework. The scalar field gives
rise to an equivalent cosmological constant that decays according to a power
law. This offers an explanation of dark energy without the usual "fine-tuning"
problem. In this paper, we study an extension of the model that might be
relevant to inflation.

There is now experimental support for a vacuum scalar field, with the
discovery of the Higgs boson of the standard model of particle theory [3].
Extension of the standard model to grand unified theories require additional
vacuum scalar fields. Here we consider a generic scalar field $\phi$, without
inquiring into its place in particle theory. What is important for us is
renormalization, a distinctive feature of quantum field theory. In contrast to
a classical field, a quantum field has virtual processes with an unbounded
momentum spectrum. The high tail of this spectrum must be cut off at some
value $\Lambda_{0},$ for otherwise calculated physical quantities would
diverge. Furthermore, the high momentum tail cannot accurately describe the
physical system being modeled. In a self-contained field theory, this cutoff
momentum provides the only scale in theory. In physical applications, we
usually deal with processes with momenta $\Lambda<<$ $\Lambda_{0}$, and it is
useful to "hide" the degrees of freedom between $\Lambda_{0}$ and $\Lambda$,
so that $\Lambda$ becomes an effective cutoff. This is achieved through
"renormalization", a readjustment of interaction parameters. For a scalar
field, renormalization makes the self-interacting potential $V\left(
\phi,\Lambda\right)  $ dependent on $\Lambda$.

If we suppose that the universe emerges from nothing at the big bang, we would
have to conclude that the scalar potential was zero at the big bang. When we
trace back towards the big bang, we should have $V\left(  \phi,\Lambda\right)
\rightarrow0,$ when the length scale goes to zero, and $\Lambda\rightarrow
\infty$. That is, the potential should be "asymptotically free", and this
uniquely determines it as the non-polynomial Halpern-Huang potential used in
I. This potential also exhibits spontaneous symmetry breaking, i.e., the field
potential can have a minimum at $\phi\neq0$. This enables us to neglect
quantum fluctuations about the average field, which is treated classically. Of
course, quantum effects are still present through the $\Lambda$ dependence of
$V\left(  \phi,\Lambda\right)  $.

We use the RW (Robertson-Walker) metric, in which all coordinates co-expand
with length scale $a$. The scalar field must share this scale, and thus
\begin{equation}
\Lambda=\frac{\hbar}{a}%
\end{equation}
This relation sets up a dynamic feedback loop between quantum field and
gravity, because the field potential depends on $\Lambda$. As shown in I, this
dynamics makes the Hubble parameter decay in time according to a power law:%
\begin{equation}
H\sim t^{-p\text{ }}\text{ \ }\left(  0<p<1\right)  \label{power}%
\end{equation}
This implies $a\sim\exp t^{1-p}$, which signifies an accelerated expansion of
the universe, indicating "dark energy" without the fine-tuning problem. For
example, with $p\approx1$, \ we can set $H=1$ initially (in Planck units), and
it will decay to the present value of $10^{-60}$ in the course of $10^{10}$
years [4].

In paper I, we had considered two questions related to cosmic inflation:

\begin{itemize}
\item What mechanism is responsible for the creation of all the matter in the
universe before the cosmic inflation made them fall out of each other's horizon?

\item How could the matter energy scale, which is of order 1 GeV, be decoupled
from the Planck scale of 10$^{18}$ GeV that is built into Einstein's equation?
\end{itemize}

With the real scalar field used in I, we were not able to find satisfactory
answers. In this paper, we extend the model to a complex scalar field, and
exploit new physics arising from the phase dynamics of the scalar field:
superfluidity, quantized vorticity, and quantum turbulence.

A complex scalar field is an order parameter for superfluidity, whatever its
microscopic origin. The point is that the gradient of its phase gives rise to
a superfluid velocity field, as is familiar from superconductivity [5], and
superfluidity in liquid helium [6,7] and cold trapped atomic gases [8]. With a
complex scalar field permeating all space, the whole universe becomes a superfluid.

To describe the cosmic superfluidity in a mathematically tractable way, it is
important to stay within the FLRW framework, for otherwise we will have to
devise a new metric --- a formidable task, considering the fact that, in
nearly a century of general relativity, we only know about a handful of
metrics. But we face an immediate problem, namely, FLRW assumes spatial
uniformity, but a completely uniform complex scalar field is not qualitatively
different from a real field, which we have already found to be inadequate. The
solution is to consider a complex scalar field with uniform modulus but
variable phase, which should still be uniform in some sense. This leads us to
consider a uniform superfluid with a uniform distribution of quantized vortices.

A quantized vortex refers to a velocity field flowing about a axis, the vortex
line, with quantized circulation. The vortex line forms a closed curve in
space, and the superfluid density goes to zero at the line. To stay in the
FLRW framework, we encase the vortex line in a tube of finite radius, and take
the superfluid density to be zero inside this tube, and uniform outside. Our
scalar field is now real and uniform, but it exists in a multiply-connected
space that is laced with "worm holes"--- vortex tubes of fluctuating sizes and
shapes. From a coarse-grained point of view, we describe the latter in terms
of a vortex-line\ density (length per unit volume), and take the density to be
uniform in space.

The uniform vortex density corresponds to what Feynman [9] calls a "vortex
tangle", or quantum turbulence. Its time development can be describe by
Vinen's equation [10], which was proposed for liquid helium on a
phenomenological basis, and subsequently derived by Schwarz [11,12] from
superfluid hydrodynamics. We adopt this equation as part of our cosmological
equations, thus picturing the early universe as dominated by quantum
turbulence in the cosmic superfluid. This is the new physics that provides
answers to the questions we raised on inflation.

To maintain a vortex tangle, there must be a supply of large vortex rings
generated continually by heat currents in the superfluid. These large rings
will degrade into ever smaller rings through reconnection of vortex lines,
until they reach microscopic size and disappear into background fluctuations.
When the supply of large rings become inadequate, the vortex tangle will decay.

The signature of a reconnection event is the creation of two oppositely
directed jets of energy, which provides an efficient mechanism for matter
creation. In this process, two originally smooth vortex lines cross, reconnect
and separate, leaving cusps in the separating lines.\ Because of the cusps,
the separating lines spring away from each other, theoretically with infinite speed
(See Appendix B), creating energy jets that could materialize via the coupling
between the superfluid and matter. By dimensional analysis, reconnections
should occur at the rate of about one per Planck volume ($10^{-99}$ cm$^{3}$)
per Planck time (10$^{-43}$s), with each reconnection releasing the order of
one unit of Planck energy (10$^{18}$ GeV). Even though the vortex tangle may
have a very short lifetime, an amount of energy of the order of the present
total energy of the universe can be released, when we factor in the total
expanding volume of the universe. This conceptually answers our quest for an
efficient way to create matter.

It is interesting to note that solar flares on the sun's surface convert a
large amount of potential energy to kinetic energy in a very short time, and
this is achieved through the reconnection of magnetic flux lines, and the
energy is released in the form of jets.

On the basis of the physical ideas described above, we put together a closed
set of phenomenological cosmological equations, in which the independent
degree of freedom are the gravitational field, the modulus of the scalar
field, the vortex-line density, and matter described as a classical perfect
fluid. This generalizes the model discussed in I.

The vortex-matter interaction brings into our equations a new scale, the QCD
scale of 1 GeV, as compared to the Planck scale of 10$^{18}$ GeV in Einstein's
equation. This two scales decouple from each other, due to the structure of
Vinen's equation. The set of cosmological equations can be split into two
subsets, one governing the gravitational and scalar fields, the other
describing the vortex-matter system, with a link whose time rate of change 
depends on the
ratio (matter scale)/(Planck scale) $\sim10^{-18}$. Decoupling occurs dues to
the extreme smallness of this number. This explains, from the viewpoint of
Einstein's equation, why one can do calculations on stellar structure\ without
having to worry about cosmic expansion, and vice versa.

The cosmological equations with appropriate initial conditions lead to a rapid
expansion of the universe. A vortex tangle initially grows, and eventually
decays, but not before it creates all the matter in the universe. In our
picture, the inflation era is the era of quantum turbulence. We can choose
phenomenological parameters such that the lifetime of the vortex tangle is of
order 10$^{-26}$s, during which time the radius of the universe increases by a
factor of order 10$^{27}$, and the total amount of matter created was equal to
what we have now, of the order of $10^{22}$ suns.

With the demise of the vortex tangle, the inflation era ends, and our model
ceases to be valid, because the universe would have grown to be so large that
density variations become important. However, with the decoupling of scales,
the model prepares the stage for the standard "hot big bang theory" [13], with
the added feature that the universe remains a superfluid. All astrophysical
activities therefore take place in the cosmic superfluid, with observable
consequences. Among these are dark matter, the galactic voids, the so-called
"non-thermal filaments", and cosmic jets. These will be discussed later in
this paper.

In summary, this model offers explanations of diverse cosmic phenomena from a
unified picture, namely a cosmic superfluid arising from a vacuum complex
scalar field, with a full range of vortex activities.

\section{Complex scalar field and superfluid vortex dynamics}

A complex scalar field $\phi\left(  x\right)  $ is equivalent to a
two-component real field $\{\phi_{1}\left(  x\right)  ,\phi_{2}\left(
x\right)  \}$, with the relation%
\begin{align}
\phi &  =\frac{\phi_{1}+i\phi_{2}}{\sqrt{2}}=Fe^{i\sigma}\nonumber\\
\phi^{\ast}  &  =\frac{\phi_{1}-i\phi_{2}}{\sqrt{2}}=Fe^{-i\sigma}%
\end{align}
where we introduce the phase representation, with modulus $F$ and phase
$\sigma.$The classical Lagrangian density is given by%
\begin{equation}
\mathcal{L}_{\phi}=-g^{\mu\nu}\partial_{\mu}\phi^{\ast}\partial_{\nu}\phi-V
\end{equation}
In the quantum field theory, in order to tame high-frequency virtual
processes, the operator $g^{\mu\nu}\partial_{\mu}\partial_{\nu}$ in the
kinetic term is "regulated" by the introduction of a small-distance cutoff, or
equivalently a high-momentum cutoff $\Lambda$. As in I, we set
$\Lambda=a^{-1}\left(  t\right)  \,$, where $a(t)$ is the scale of the RW metric.\ 

We use the Halpern-Huang (HH) potential%
\begin{align}
V\left(  \phi\right)   &  =\Lambda^{4}U_{b}(z)\nonumber\\
U_{b}(z)  &  =ca^{b}\left[  M\left(  -2+b/2,1,z\right)  -1\right] \nonumber\\
z  &  =16\pi^{2}\Lambda^{-2}\phi^{\ast}\phi
\end{align}
Its derivative can be represented in the form%
\begin{align}
\frac{\partial V}{\partial\phi^{\ast}}  &  =\Lambda^{4}U_{b}^{\prime}%
(z)\frac{dz}{d\phi^{\ast}}=\Lambda^{4}16\pi^{2}\phi U_{b}^{\prime
}(z)\nonumber\\
U_{b}^{\prime}(z)  &  =-c\Lambda^{-b}\left(  2-\frac{b}{2}\right)  M\left(
-1+b/2,2,z\right)
\end{align}
The classical equation of motion of the scalar field is%
\begin{equation}
\partial_{\mu}\left[  \sqrt{-g}g^{\mu\nu}\partial_{\nu}\phi\right]  -\sqrt
{-g}\frac{\partial V}{\partial\phi^{\ast}}=0
\end{equation}
In the phase representation this reads%
\begin{align}
\frac{1}{\sqrt{-g}}\partial_{\mu}\left(  \sqrt{-g}g^{\mu\nu}\partial_{\nu
}F\right)  -g^{\mu\nu}F\partial_{\mu}\sigma\partial_{\nu}\sigma-\frac{1}%
{2}\frac{\partial V}{\partial F}  &  =0\nonumber\\
\frac{1}{\sqrt{-g}}\partial_{\mu}\left(  \sqrt{-g}g^{\mu\nu}\partial_{\nu
}\sigma\right)   &  =0 \label{field0}%
\end{align}

The complex scalar field is commonly used in condensed matter physics as the
order parameter for superfluidity, with the superfluid velocity defined by%

\begin{equation}
\mathbf{v}=\nabla\sigma
\end{equation}
This is of dimension $\left(  \text{length}\right)  ^{-1}$. To obtain a
velocity, it is customary to multiply $\nabla\sigma$ by a unit of vorticity
$\kappa_{0}=h/m_{0}$, where $h$ is Planck's constant, and $m_{0}$ is a mass
parameter. For simplicity we omit this factor.

The presence of a vacuum complex scalar field makes the universe a superfluid,
a salient feature of which is the quantization of vorticity. Around any closed
circuit $C$, the phase $\sigma$ can only change by a multiple of $2\pi,$ since
$\phi\left(  x\right)  $ must be continuous. This lead to the quantization
condition
\begin{equation}%
%TCIMACRO{\doint \limits_{C}}%
%BeginExpansion
{\displaystyle\oint\limits_{C}}
%EndExpansion
\mathbf{v}\cdot d\mathbf{s=}2\pi n
\end{equation}
where the line integral is carried around any closed curve $C$ in space, and
$n$ is an integer. If $n\neq0$, then $C$ cannot be shrunken to zero; it
encircles a directed line called the vortex line, on which $\phi=0$. The
vortex line cannot terminate inside the superfluid; it either forms a closed
loop, or terminate on a surface. We only need consider $n=1$, for higher
vortices tend to be unstable and break up into lower ones, when perturbations
are present.

The velocity tends to infinity at the vortex line, and the modulus $F$ must
vanish on the line to keep the energy finite. Thus, the vortex line renders
the space non-simply connected, and we can have $\nabla\times\mathbf{v}\neq0$,
even though $\mathbf{v}$ is a gradient. We write%
\begin{equation}
\nabla\times\mathbf{v=j}%
\end{equation}
where $\mathbf{j}\left(  x\right)  $ is the vorticity density. This is
analogous to Maxwell's equation for a magnetic field due to a current in a
wire shaped like\ the vortex line. The solution is the Biot-Savart law%
\begin{equation}
\mathbf{v}\left(  \mathbf{r},t\right)  =\int_{\mathcal{L}}\frac{\left(
\mathbf{s}-\mathbf{r}\right)  \times d\mathbf{s}}{\left\vert \mathbf{s}%
-\mathbf{r}\right\vert ^{3}}%
\end{equation}
where $\mathbf{s}$ is the vector position of a point on the vortex line, and
the integration ranges over $\mathcal{L}$, the totality of all vortex lines.
The integral above diverges when $\mathbf{s}\rightarrow\mathbf{r}$, and a
cutoff is needed. In liquid helium the cutoff comes from atomicity, and in our
case it comes from the field-theory cutoff $\Lambda^{-1}=a\left(  t\right)  $.
This replaces the vortex line by a tube called the vortex core, whose radius
should be proportional to $a\left(  t\right)  $, since that is the only length
scale available. We shall continue to refer to the center of the core as the
vortex line.

The field modulus $F$ inside the vortex core is suppressed, with functional
form determined by the cutoff function. We adopt the simple model that $F$ is
zero inside, and constant outside. In this picture, the scalar field can be
regarded as uniform in space, except that the space is made non
simply-connected, by exclusion of the vortex tube. A static vortex solution is
discussed in Appendix A, and vortex dynamics is reviewed in Appendix B.

The vortex tube cannot spontaneously arise, but must be nucleated by quantum
fluctuations of the scalar field. A microscopic ring-shape tube would appear
by fluctuation, and grow to macroscopic dimensions under appropriate
conditions. The vortex tube has an energy cost of $\epsilon_{0}$ per unit
length, and its presence induces superfluid flow. Thus, its energy consists of
of two parts:%
\begin{align}
\epsilon_{0}  &  =\text{energy per unit length of vortex tube}\nonumber\\
v^{2}  &  =\text{energy density of induced superfluid flow}%
\end{align}

We work within the RW metric, which assumes spatial uniformity. To conform to
this requirement in the equations of motion (\ref{field0}), we assume that $F$
is constant in space outside of the vortex tube, and we perform spatial
averages on terms involving the phase $\sigma$. The resulting equations of
motion are
\begin{align}
\ddot{F}  &  =-3H\dot{F}+F\left\langle \dot{\sigma}^{2}\right\rangle
-F\left\langle \left\vert \nabla\sigma\right\vert ^{2}\right\rangle -\frac
{1}{2}\frac{\partial V}{\partial F}\nonumber\\
\frac{d}{dt}\left\langle \dot{\sigma}\right\rangle  &  =-3H\left\langle
\dot{\sigma}\right\rangle \label{field1}%
\end{align}
where $H=\dot{a}/a$, and $\left\langle {}\right\rangle $ denotes spatial
average. The space, however, is the non-simply connected region outside of
vortex tubes. The energy density and pressure of the scalar field are given by%
\begin{align}
\rho_{\phi}  &  =\dot{F}^{2}+\left\langle \dot{\sigma}^{2}\right\rangle
+V\nonumber\\
p_{\phi}  &  =\dot{F}^{2}+\left\langle \dot{\sigma}^{2}\right\rangle
-V-\frac{a}{3}\frac{\partial V}{\partial a}%
\end{align}
where the $\partial V/\partial a$ term is explained in I. The second equation
in (\ref{field1}) gives $\left\langle \dot{\sigma}\right\rangle \varpropto
a^{-3},$which will rapidly vanish as $a$ increases. We assume $\left\langle
\dot{\sigma}^{2}\right\rangle \sim O\left(  a^{-6}\right)  $, and neglect it.
Thus we have%
\begin{equation}
\ddot{F}=-3H\dot{F}-F\left\langle v^{2}\right\rangle -\frac{1}{2}%
\frac{\partial V}{\partial F} \label{scalar1}%
\end{equation}
with%
\begin{align}
\rho_{\phi}  &  =\dot{F}^{2}+V+\left\langle v^{2}\right\rangle \nonumber\\
p_{\phi}  &  =\dot{F}^{2}-V-\left\langle v^{2}\right\rangle -\frac{a}{3}%
\frac{\partial V}{\partial a}%
\end{align}

The vortex tubes created in the early universe must have a core radius
proportional to $a\left(  t\right)  $ of the RW metric, since that is only
length scale available. This core will expand with the universe, maintaining
the same fraction of the radius of the universe, and may account for the
presently observed voids in the galactic distribution, as we shall discuss later.

\section{Vinen's equation}

The formation of quantum turbulence in the form of a vortex tangle is
discussed in Appendix B. In this model, which is restricted to spatial uniformity
through use of the RW metric, we describe the tangle with one variable
$\ell\left(  t\right)  $, the vortex line density (average length per unit
volume). This quantity obeys Vinen's phenomenological equation, which in flat
space-time has the form
\begin{equation}
\dot{\ell}=A\ell^{3/2}-B\ell^{2}%
\end{equation}
where $A$ and $B$ are phenomenological parameters. The generalization to
curved space-time is
\begin{equation}
g^{-1/2}\frac{d}{dt}\left(  g^{1/2}\ell\right)  =A\ell^{3/2}-B\ell^{2}%
\end{equation}
which in RW metric reduces to%
\begin{equation}
\dot{\ell}=-3H\ell+A\ell^{3/2}-B\ell^{2}%
\end{equation}
The energy density of the vortex tangle is%
\begin{equation}
\rho_{\text{v}}=\epsilon_{0}\ell
\end{equation}
Vinen's equation thus states%
\begin{equation}
\dot{\rho}_{\text{v}}=-3H\rho_{\text{v}}+\alpha\rho_{\text{v}}^{3/2}-\beta
\rho_{\text{v}}^{2} \label{vinen}%
\end{equation}
where $\alpha$ and $\beta$ are model parameters that may depend on the time.

As explained in Appendix B, two vortex lines undergo reconnection when they
approach each other to within a distance $\delta\propto$ $v^{-1}$, where $v$
is their relative speed, which is of the same order as the average speed in
the superfluid. Thus, in steady-state, the average spacing between vortex
lines should be $\delta$. On the other hand, by geometrical considerations,
the average spacing should be of order $\ell^{-1/2}.$ This gives the estimate%
\begin{equation}
\left\langle v^{2}\right\rangle =\zeta_{0}\rho_{\text{v}} \label{zeta}%
\end{equation}
where $\zeta_{0}$ is a constant.

The parameters $\alpha,\beta,\zeta_{0}$ may depend on $a\left(  t\right)  $,
for they could depend on the radius of the vortex core .

\section{Cosmological equations with quantum turbulence and matter creation}

Let us review the framework for the cosmological equations, i.e., Einstein's
equation with RW metric. The cosmic expansion is described by $a(t)$, the
scale of the RW metric, and we introduce the Hubble parameter $H=\dot{a}/a$.
The equation are (with $4\pi G=1)$%

\begin{align}
\dot{H}  &  =\frac{k}{a^{2}}-\left(  p+\rho\right) \nonumber\\
X  &  \equiv H^{2}+\frac{k}{a^{2}}-\frac{2}{3}\rho=0\nonumber\\
\dot{\rho}  &  =3H\left(  \rho+p\right)
\end{align}
where $\rho$ and $p$ are respectively the total energy density and pressure
derived from the energy-momentum tensor $T^{\mu\nu}$\ of non-gravitational
systems. The second equation, of the form $\dot{X}=0,$ is a constraint on
initial values. The third equation is the conservation law $T_{;\mu}^{\mu\nu
}=0$, and it guarantees $\dot{X}=0$. The inclusion of (\ref{scalar1}) and
(\ref{vinen}) will complete the dynamics and close the equations.

So far we have three independent variables: the scale of the universe $a$, the
modulus of the vacuum scalar field $F$, and the energy density of the vortex
tangle $\rho_{\text{v}}$. We now introduce matter, modeled as a classical
perfect fluid of energy density $\rho_{\text{m}}$. Its pressure is taken to be
$p_{\text{m}}=w_{0}\rho_{\text{m}}$, where $w_{0}$ is the equation-of-state
parameter, with possible values $\{-1,0,1/3\}$ corresponding respectively to
"vacuum energy", "pressureless dust", and "radiation". The total energy
density $\rho$ and total pressure $p$ are now given by
\begin{align}
\rho &  =\rho_{\phi}+\rho_{\text{m}}+\rho_{\text{v}}\nonumber\\
p  &  =p_{\phi}+w_{0}\rho_{\text{m}}%
\end{align}
The cosmological equations are then given by%
\begin{align}
\dot{H}  &  =\frac{k}{a^{2}}-\left(  \rho+p\right) \nonumber\\
\ddot{F}  &  =-3H\dot{F}-F\left\langle v^{2}\right\rangle -\frac{1}{2}%
\frac{\partial V}{\partial F}\nonumber\\
\dot{\rho}_{\text{v}}  &  =-3H\rho_{\text{v}}+\alpha\rho_{\text{v}}%
^{3/2}-\beta\rho_{\text{v}}^{2}%
\end{align}
with constraint and conservation equations%

\begin{align}
X  &  \equiv H^{2}+\frac{k}{a^{2}}-\frac{2}{3}\rho=0\nonumber\\
\dot{X}  &  =0
\end{align}

The equation $\dot{X}=0$ acts as the equation of motion for matter. To
emphasize this, we can rewrite the cosmological equations in the following form:%

\begin{align}
\dot{H}  &  =\frac{k}{a^{2}}-2\dot{F}^{2}+\frac{a}{3}\frac{\partial
V}{\partial a}-\left(  1+w_{0}\right)  \rho_{\text{m}}-\rho_{\text{v}%
}\nonumber\\
\ddot{F}  &  =-3H\dot{F}-\zeta_{0}\rho_{\text{v}}F-\frac{1}{2}\frac{\partial
V}{\partial F}\nonumber\\
\dot{\rho}_{\text{v}}  &  =-3H\rho_{\text{v}}+\alpha\rho_{\text{v}}%
^{3/2}-\beta\rho_{\text{v}}^{2}\nonumber\\
\dot{\rho}_{\text{m}}  &  =-3H\left(  1+w_{0}\right)  \rho_{\text{m}}%
-\alpha\rho_{\text{v}}^{3/2}+\beta\rho_{\text{v}}^{2}+\frac{dF^{2}}{dt}%
\zeta_{0}\rho_{\text{v}} \label{cosmo0}%
\end{align}
where the last equation is a rewrite of $\dot{X}=0.$ The parameters
$\alpha,\beta$ are defined in (\ref{vinen}), and may depend on the time. The
constraint on initial conditions%
\begin{equation}
X\equiv H^{2}+\frac{k}{a^{2}}-\frac{2}{3}\rho=0
\end{equation}
is now preserved by the equations of motion.

Finally, we introduce the total energies
\begin{align}
E_{\text{v}}  &  =a^{3}\rho_{\text{v}}\nonumber\\
E_{\text{m}}  &  =a^{3\left(  1+w_{0}\right)  }\rho_{\text{m}}%
\end{align}
which will absorb the kinematic terms proportional to $3H$ in the equations.
Note that $w_{0\text{ }}$appears above as an ``anomalous dimension" [14]. For
simplicity, we put $w_{0}=0$, corresponding to pressureless dust. The
cosmological equations plus constraint then become
\begin{align}
\dot{H}  &  =\frac{k}{a^{2}}-2\dot{F}^{2}+\frac{a}{3}\frac{\partial
V}{\partial a}-\frac{1}{a^{3}}\left(  E_{\text{m}}+E_{\text{v}}\right)
\nonumber\\
\ddot{F}  &  =-3H\dot{F}-\frac{\zeta_{0}}{a^{3}}E_{\text{v}}F-\frac{1}{2}%
\frac{\partial V}{\partial F}\nonumber\\
\dot{E}_{\text{v}}  &  =s_{1}E_{\text{v}}^{3/2}-s_{2}E_{\text{v}}%
^{2}\nonumber\\
\dot{E}_{\text{m}}  &  =-s_{1}E_{\text{v}}^{3/2}+s_{2}E_{\text{v}}^{2}%
+\frac{dF^{2}}{dt}\zeta_{0}E_{\text{v}}\nonumber\\
X  &  \equiv H^{2}+\frac{k}{a^{2}}-\frac{2}{3}\rho=0 \label{cosmoequ}%
\end{align}
where%
\begin{equation}
\rho=\dot{F}^{2}+V+\frac{1+\varsigma_{0}}{a^{3}}E_{\text{v}}+\frac{1}{a^{3}%
}E_{\text{m}}%
\end{equation}
and%
\begin{align}
s_{1}  &  =\alpha a^{-3/2}\nonumber\\
s_{2}  &  =\beta a^{-3}%
\end{align}
where $\zeta_{0}$ is a constant defined in (\ref{zeta}), and $\alpha,\beta$
are time-dependent parameters defined in (\ref{vinen}). This constitutes a
self-consistent and self-contained initial-value problem.

\begin{figure}
[ptb]
\begin{center}
\includegraphics[
%natheight=1.039500in,
%natwidth=1.209900in,
%height=1.919in,
width=3in
]%
{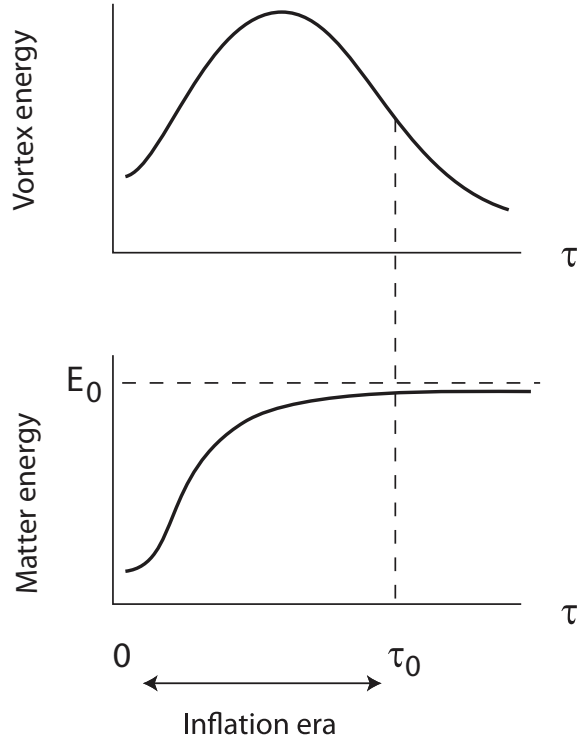}%
\caption{Upper panel shows total energy of the vortex tangle (quantum
turbulence) as function of nuclear time $\tau$, which is related to the Planck
time $t$ by $\tau=s_{1}t$, with $s_{1}\sim10^{-18}$. The lifetime $\tau_{0}$
of the vortex tangle is the duration of the inflation era, which can be
estimated to be $10^{-26}$s. By the same estimate, the radius of the universe
increased by a factor 10$^{27}$. Lower panel show total energy of matter
produced, which is proportional to the area under the curve in the upper
panel. The total energy $E_{0}$ can be adjusted to correspond to the total
obsserved energy in the universe, the order of 10$^{22}$ solar masses.}%
\end{center}
\end{figure}

\section{Decoupling}

The vortex-matter system as described by $E_{\text{m}}$ and $E_{\text{v}}$
should be governed by a QCD energy scale of about 1 GeV, as compared to the
Planck scale of 10$^{18}$ GeV. The QCD scale is unrelated to the quark mass
from the Higgs field, but spontaneously arises in the scale-invariant QCD,
through formation of the nucleon bound state, in a phenomenon known as
"dimensional transmutation". The simplest mathematical example of this
mechanism is the occurrence of a bound state in an attractive $\delta
$-function potential in the 2D Schr\"{o}dinger equation [15].

We incorporate the new scale into the equation by asserting that $s_{1}%
,s_{2},$ which are measured in Planck units, to be order $10^{-18}$. That is,
we put%
\begin{align}
s_{1}  &  =\mu\kappa_{1}\nonumber\\
s_{2}  &  =\mu\kappa_{2} \label{kappa}%
\end{align}
where%
\begin{equation}
\mu=\frac{\text{Planck time scale}}{\text{Nuclear time scale}}=\frac
{\text{Nuclear energy scale}}{\text{Planck energy scale}}\sim10^{-18}%
\end{equation}
and $\kappa_{1},\kappa_{2}$ are time-dependent parameters of order unity. The
last two cosmological equations can be rewritten in the form%
\begin{align}
\frac{dE_{\text{v}}}{d\tau}  &  =\kappa_{1}E_{\text{v}}^{3/2}-\kappa
_{2}E_{\text{v}}^{2}\nonumber\\
\frac{dE_{\text{m}}}{d\tau}  &  =-\kappa_{1}E_{\text{v}}^{3/2}+\kappa
_{2}E_{\text{v}}^{2}+\frac{\zeta_{0}}{\mu}\frac{dF^{2}}{dt}E_{\text{v}}
\label{vm}%
\end{align}
where%
\begin{equation}
\tau=\mu t
\end{equation}
The only link to the first two cosmological equations is the quantity
$\zeta_{0}\mu^{-1}dF^{2}/dt$, which is extremely rapidly varying in terms of
$\tau$. We can average over $t$, obtaining a function of $\tau:$
\begin{equation}
K_{0}\left(  \tau\right)  =\frac{\zeta_{0}}{\mu}\left\langle \frac{dF^{2}}%
{dt}\right\rangle _{t}%
\end{equation}
Because of the factor $\mu^{-1}\sim10^{18}$, this gives an "anomalously large"
rate of matter production.

The cosmological equations (\ref{cosmoequ}) breaks up into two sets governed
by different time scales. The last two are governed by the matter time scale,
as discussed above, and the first two equations are governed by the Planck
scale. The vortex-matter system affects cosmic expansion through the quantity
$E_{\text{m}}+E_{\text{v}}$ , which is practically a constant with respect to
the Planck scale. Treating it as a constant will yield qualitatively the same
as that in I, the power-law (\ref{power}).

In summary, decoupling happens because

\begin{itemize}
\item From the viewpoint of the cosmic expansion, the vortex-matter system is
essentially static.

\item From the viewpoint of the vortex-matter system, the cosmic expansion is
extremely fast, but it is noticeable only as an abnormally large rate of
matter production.
\end{itemize}

\begin{figure}
[ptb]
\begin{center}
\includegraphics[
%natheight=1.039500in,
%natwidth=1.209900in,
%height=1.919in,
width=5.6in
]%
{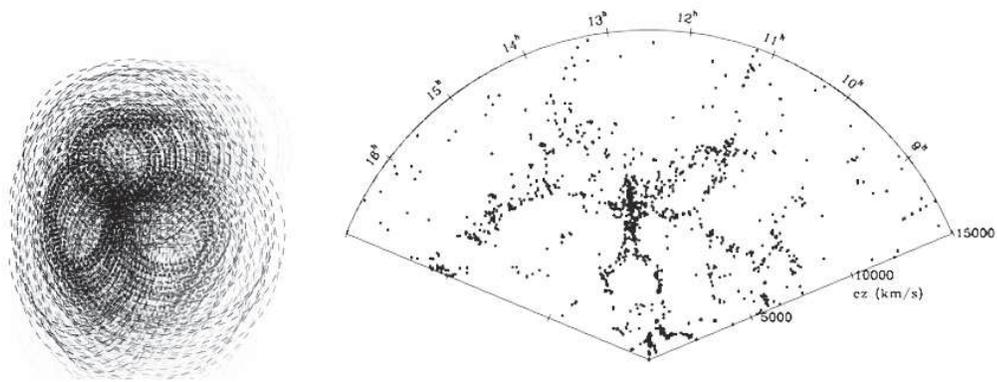}%
\caption{Left panel: Simulation of galactic voids by superposition of three
vortex tubes, whose cores, originally of Planck scale near the big bang, have
grown with the expanding universe, and reached hundreds of million of light
years, in the 15 billion years since. The vortex cores are devoid of he vacuum
scalar field, and therefore of matter. Galaxies formed outside adhere to tube
surfaces due to hydrodynamic pressure. Right panel: The "stickman"
configuration observed in galactic distributions, from Ref.[18].}%
\end{center}
\end{figure}

\section{Inflation}

The inflation scenario is designed to explain the presently observed
large-scale uniformity of galactic distribution in the universe. It assumes
that all matter was created when the universe was so small that they stay
within each other's event horizon, and so maintain a uniform density. The era
comes to an end when, with the expansion of the universe, its size is inflated
to such an extend that the matter fall out of each other's event horizon, but
they retain the memory of a uniform density. Traditional estimates puts the
inflation factor at some 27 orders of magnitude [13].

It will take more study to determine the correct choice of the
phenomenological functions $\kappa_{1},\kappa_{2}$ in (\ref{vm}). Apart from
physical considerations, one faces the technical problem of numerical solution
of the cosmoogical equations with two vastly different scales. We leave
detailed studies for the future.

Here we make some postulates and simplifications, in order to illustrate how
inflation can happen in this model. We decouple the equation "by hand", by
setting $K_{0}\left(  \tau\right)  $ to be a large constant, and choose%

\begin{align}
\kappa_{1}  &  =\frac{A}{1+B\tau}\nonumber\\
\kappa_{2}  &  =1
\end{align}
where $A,B$ are constants. The physical reasoning is that $\kappa_{1}$
describes the "head wind" needed to fuel the growth of quantum turbulence, and
it decreases with $\tau$ because of cosmic expansion.

In the second equation of (\ref{vm}), we can neglect the first two terms,
because $K_{0}$ is large. Then we have%

\begin{align}
\frac{dE_{\text{v}}}{d\tau}  &  =-E_{\text{v}}^{2}+\frac{A}{1+B\tau
}E_{\text{v}}^{3/2}\nonumber\\
\frac{dE_{\text{m}}}{d\tau}  &  =K_{0}E_{\text{v}} \label{v-m}%
\end{align}
These can be regarded as phenomenological equation for inflation. The
qualitative behavior of the solution, which can be seen by inspection, is
illustrated in Fig.1. The vortex energy $E_{\text{v}}$ rises through a maximum
and decays with a long tail, like $\tau^{-1}$. The characteristic time
$\tau_{0}$ defines the lifetime of quantum turbulence, and therefore that of
the inflation era. The total matter energy $E_{\text{m}}$ is proportional to
the area under the curve for $E_{\text{v}}$. It approaches a constant $E_{0}$,
which is the total energy of matter created during the inflation era.

We now put in some numbers. The lifetime of the tangle $\tau_{0}$ corresponds
to the Planck time $t_{0}=\mu^{-1}\tau_{0}$. According to the power-law
obtained in I, the radius of the universe expands by a factor $a\left(
t_{0}\right)  /a_{0}=\exp\left(  \xi t_{0}^{1-p}\right)  $. Taking $\tau
_{0}\sim1$, $\xi=1,$ $p=0.9$, we obtain%
\begin{align}
t_{0}  &  \sim10^{18}\text{ (}10^{-26}\text{s)}\\
\frac{\ a\left(  t_{0}\right)  }{a_{0}}  &  \sim10^{27}%
\end{align}
We can adjust $K_{0}$ to yield the total energy in the universe:%
\begin{equation}
E_{0}\approx10^{22}m_{\text{sun}}=2\times10^{69}\text{joule}%
\end{equation}

Our picture of inflation is completely different from the conventional one
[13]. In the latter, the scalar field starts at zero field at a "false vacuum"
corresponding to a potential maximum. It then "rolls" slowly to the potential
minimum, the "true vacuum". The field then oscillates about the minimum, and
creates matter through "reheating". In our case, as shown in I, there is no
"slow roll". Instead, the field performs rapid oscillations with very large
amplitudes, and bounces off the exponential wall of the HH potential. During
this era of rapid oscillation, a vortex tangle rises and falls, and all matter
was created in the turbulence, (physically, through the vortex reconnections
essential to the tangle's maintenance).

\begin{figure}
[ptb]
\begin{center}
\includegraphics[
%natheight=1.039500in,
%natwidth=1.209900in,
%height=1.919in,
width=5in
]%
{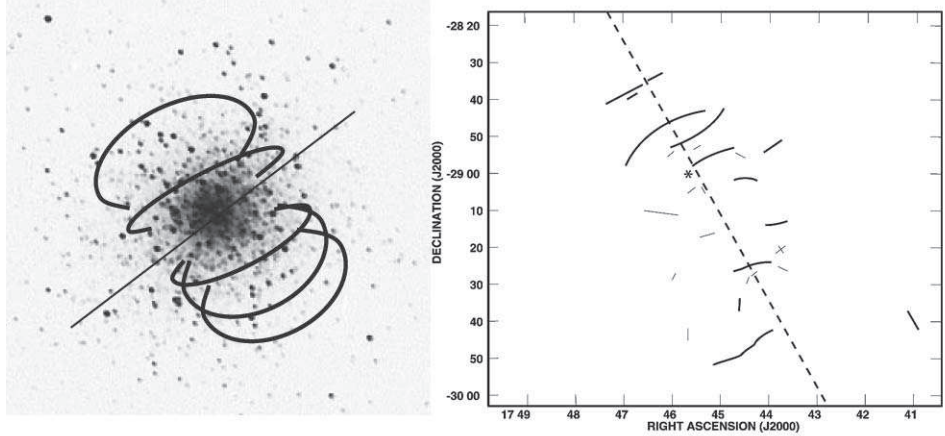}%
\caption{Left panel shows a drawing of a rotating stellar distribution, which
could drag along the cosmic superfluid it is immersed in, if it has sufficient
randomness. The stellar system will then acquire extra moment of inertia,
perceived by us as ``dark mass''. The co-moving superfluid will be separated
from the stationary background fluid by a boundary layer that is laced with
vortex tubes. These could be the ``non-thermal filaments'' observed near the
center of the Milky Way, a schematic drawing of which, from Ref.[25], is
reproduced in the right panel.}%
\end{center}
\end{figure}

We have relied on a scalar field to produce quantum turbulence. There is
suggestion that it could arise naturally in string theory [16].

\section{The post-inflation universe}

After the inflation era, the standard hot big bang scenario takes over. In
this regime, spatial non-uniformity becomes the interesting feature, and our
model ceases to be valid. It pave the way for hot big bang theory through
decoupling of nucleogenesis and galaxy formation from cosmic expansion. The
model, however, has an important legacy: it leaves the universe in a
superfluid state, and this leads to observable manifestations, as discussed in
the following.

\begin{figure}
[ptb]
\begin{center}
\includegraphics[
%natheight=1.039500in,
%natwidth=1.209900in,
%height=1.919in,
width=3in
]%
{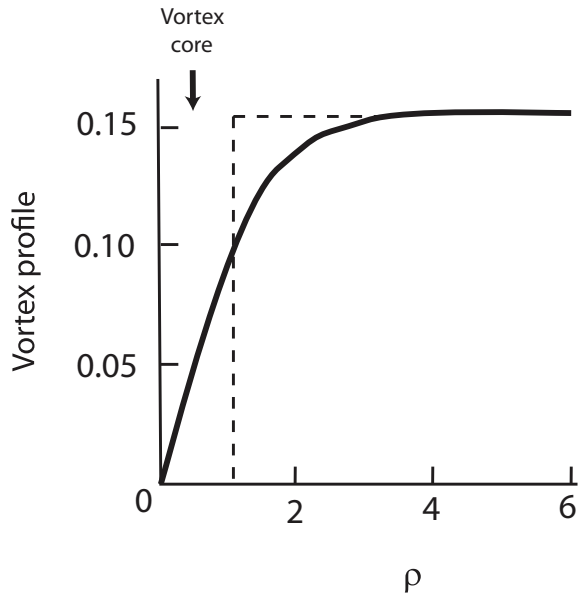}%
\caption{Profile of field modulus in a vortex solution with infinite vortex line
along the $z$-axis, as a function of reduces distance $\rho$ from the line.
The field near $\rho=0$ is further suppressed by the short-distance cutoff,
and this creates a vortex core. We approximate the configuration with a sharp
cutoff, so the the field outside the core is constant.}%
\end{center}
\end{figure}

\subsection{Galactic voids}

After the demise of the vortex tangle, there will be leftover vortex tubes.
These tubes are devoid of the scalar field, and presumably no matter was ever
created inside. Their cores must expand with the universe, and by now would
have grown to the enormous voids observed in the distribution of galaxies,
typically 10$^{8}$ light years across. Matter created outside the vortex tubes
tend to accumulate at the tube surface, due to a lowering of the hydrodynamic
pressure caused by a higher tangential superfluid velocity there. In
superfluid liquid helium, this effect has been demonstrated, through the
coating of vortex tubes by dissolved metallic nanoparticles [17]. In Fig.2 we
simulate galactic voids arising from three vortex tubes, with comparison to
the observed "stick man" configuration [18].

\subsection{ Varieties of vortices}

The galactic voids corresponds to vortex tubes in the primordial scalar field,
which were created right after the big bang. Different types of vacuum field
could emerge with the creation of matter, giving rise to different types of
superfluids with their own vortices, of different core sizes. The possible
types of vacuum fields would be determined by particle theory.

In the cosmological context, the presence of different types of superfluids
could be likened to a mixture of liquid $^{4}$He and $^{3}$He at temperatures
below 10$^{-3}$K, when both are superfluids. In such a mixture, The core of a
vortex tube could be devoid of $^{4}$He but not $^{3}$He, and vice versa, or
it could be devoid of both. Added to the complexity is the fact that the
$^{4}$He-$^{3}$He mix can exist in various phases, depending on the
temperature and the relative concentration, in which the two liquids either
commingle or segregate. In the cosmological context, the coexistence of a
variety of superfluids would present rich phenomena, on which we are not in a
position to speculate.

When we refer to "the superfluid" or "the vortex tube" in the following, we do
not commit ourselves to a specific type, but merely suggest generic behaviors.

\begin{figure}
[ptb]
\begin{center}
\includegraphics[
%natheight=1.039500in,
%natwidth=1.209900in,
%height=1.919in,
width=5in
]%
{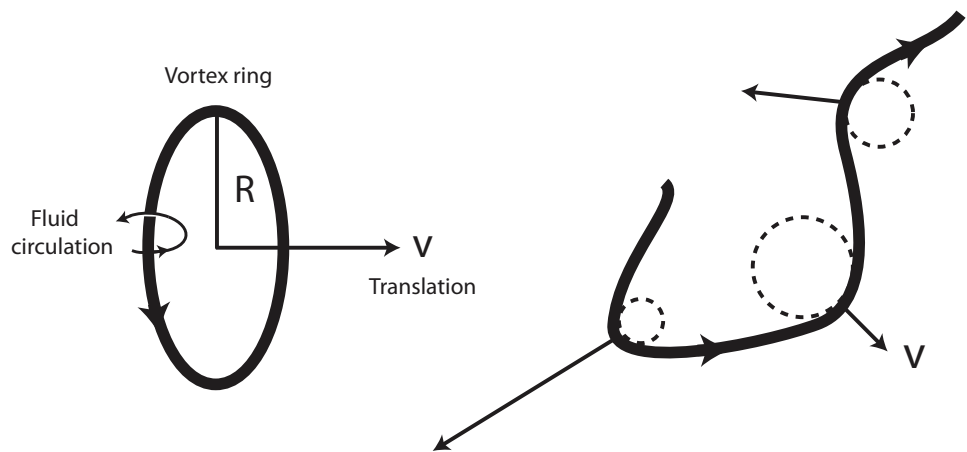}%
\caption{The heavy lines in these picture denote the vortex core, which has a
direction specified by the vorticity. The vortex ring moves in a direction
consistent with the right-hand rule, with a velocity approximately inversely
proportional to it radius. A vortex tube moves in such a fashion such that the
local velocity at any point is that of a tangential vortex ring with the local
radius of curvature.}%
\end{center}
\end{figure}

\subsection{Dark matter}

A galaxy placed in the cosmic superfluid will attract superfluid, and create a
region of superfluid density greater that in vacuum. This would be the dark
matter halo observed through gravitational lensing, as in the so-called
"bullet cluster" [19].

The galaxy can move and rotate in the superfluid halo without friction, as
long the velocity is below a critical value. Above that value, the superfluid
will rotate through the creation of quantized vortices. This may offer an
explanation of the velocity curve of galaxies [20].

However, a rough estimate shows that an average galaxy is on the margin of
critical angular velocity to create one quantized vortex, a conclusion also
reached in models of dark matter based on\ Bose-condensed particles [21,22].

There is another effect, independent of vortex formation, that may be
relevant, namely, the vacuum field can be pinned by a random potential [23].
If a galaxy is perceived by the vacuum field as a random potential, it could
drag the field along in its rotation, and acquire extra moment of inertia. The
real physics of the velocity curve of galaxies is unclear.

\subsection{Non-thermal filaments}

A fast-rotating star, such as a neutron star on its way to \ \ becoming a
black hole, would easily exceed the critical angular velocity for vortex
creation, and be encaged by vortex lines. The cores of some types of vortex
lines could trap charged matter and shine. In fact, vortex cores in liquid
helium have been made visible through the trapping of hydrogen ice [24]. In
the astrophysical context, such vortex lines could be candidates for the
"non-thermal filaments" observed near the center of the Milky Way [25], as
illustrated in Fig.3.

\subsection{Jet events}

Vortex lines in the later universe will be sparsely distributed, compared to
those in vortex tangle of the early universe; but they could find each other
occasionally and reconnect. As discuss in Appendix B, the signature of a
reconnection is the production of two jets of energy. This could be the
mechanism behind the observed gamma ray bursts and cosmic jets.

\begin{figure}
[ptb]
\begin{center}
\includegraphics[
%natheight=1.039500in,
%natwidth=1.209900in,
%height=1.919in,
width=4in
]%
{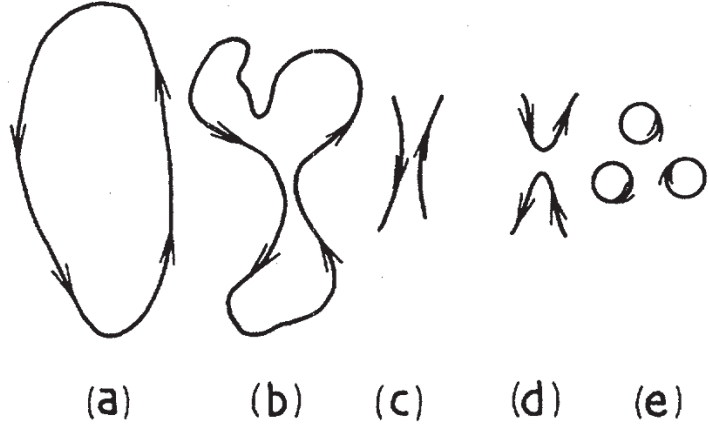}%
\caption{Feynman's sketch of the decay of a quantized vortex ring from Ref.[9].
Through reconnections, a large vortex ring become smaller rings, and smaller
rings become even smaller ones, and so on, to quantum turbulence.}%
\end{center}
\end{figure}

\begin{figure}
[ptb]
\begin{center}
\includegraphics[
%natheight=1.039500in,
%natwidth=1.209900in,
%height=1.919in,
width=5.4in
]%
{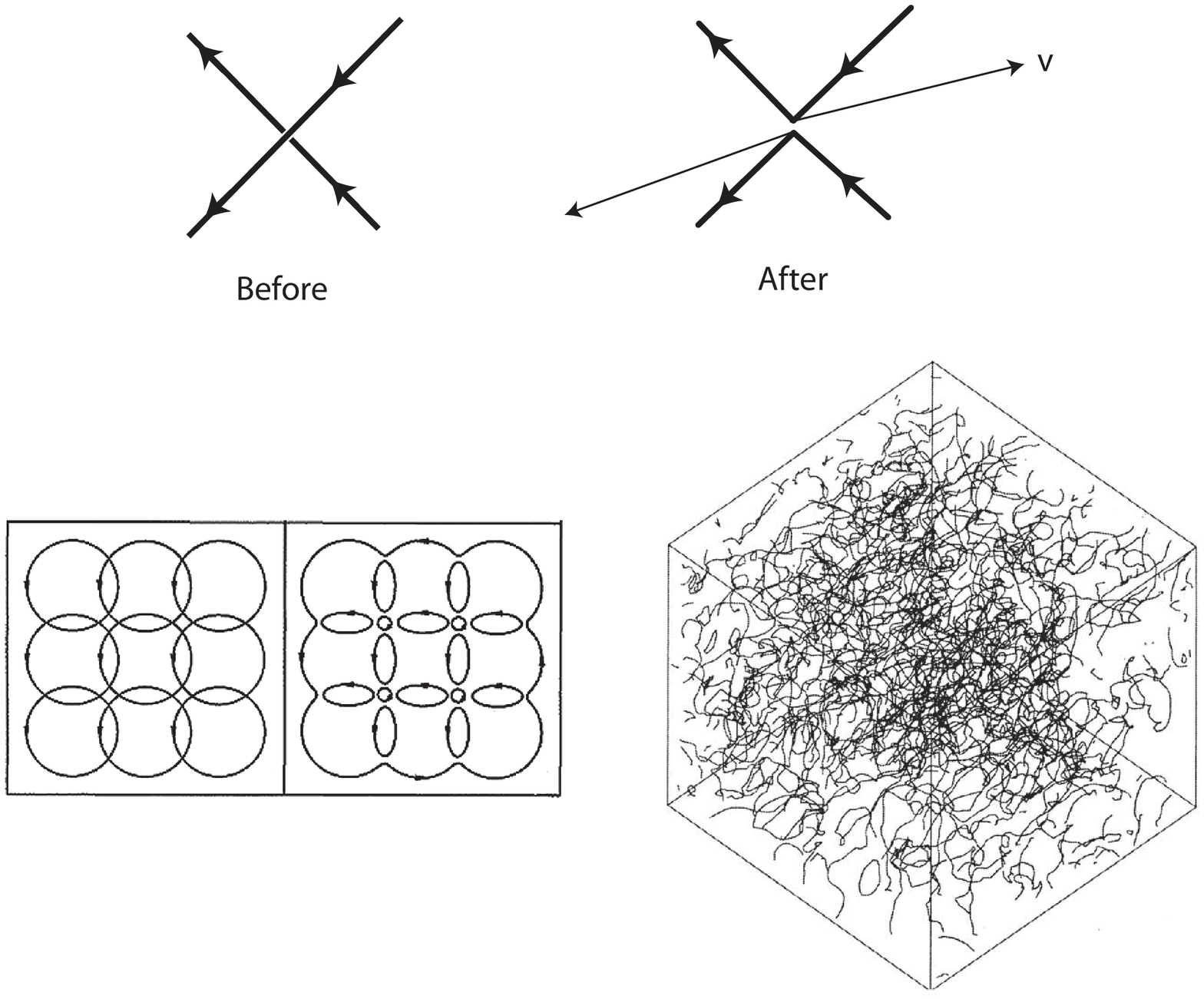}%
\caption{Upper panel: Immediately after reconnection, two cusps occur on the
participating vortex lines, which, because of the near-zero radii of
curvature, spring away from each other with theoretically infinite speed,
creating two jets of energy. Lower panel: Left side schematically illustrates
emulsification of system of vortex rings due to reconnections, from Ref.[11].
Right side show a fully-formed vortex tangle of fractal dimension 1.6., from Ref.[28].}%
\end{center}
\end{figure}

\section*{Acknowledgments}

One of us (K.H.) has benefited by conversations with D.P. Lathrop on quantum
turbulence and possible cosmological applications. In particular he thanks him
for directing attention to the \textquotedblleft non-thermal
filaments\textquotedblright.

\appendix

\section{Static vortex solution}

We solve for a static vortex solution to the complex scalar field equation in
flat space-time:%
\begin{equation}
\left(  -\frac{\partial^{2}}{\partial t^{2}}+\nabla^{2}\right)  \phi
-\frac{\partial V}{\partial\phi^{\ast}}=0
\end{equation}
where $V$ is the Halpern-Huang potential. The equations of motion in the phase
representation $\phi=Fe^{i\sigma}$are%
\begin{align}
\left(  -\frac{\partial^{2}}{\partial t^{2}}+\nabla^{2}\right)  F+F\dot
{\sigma}^{2}-F|\nabla\sigma|^{2}-\frac{\partial V}{\partial F}  &
=0\nonumber\\
\left(  -\frac{\partial^{2}}{\partial t^{2}}+\nabla^{2}\right)  \sigma
-\frac{2}{F}\frac{\partial F}{\partial t}\frac{\partial\sigma}{\partial
t}+\frac{2}{F}\nabla F\cdot\nabla\sigma &  =0
\end{align}
Consider an infinite vortex line along the $z$-axis with \ unit quantized
vorticity, such that%
\begin{equation}%
%TCIMACRO{\doint \limits_{C}}%
%BeginExpansion
{\displaystyle\oint\limits_{C}}
%EndExpansion
\nabla\sigma\cdot d\mathbf{s=}2\pi
\end{equation}
where $C$ is a circle about the origin in the $xy$ plane. This gives
$\sigma=\theta$, in cylindrical coordinates $\left(  r,\theta\right)  $. Thus
$\nabla\sigma=\hat{\theta}r^{-1}$, $|\nabla\sigma|^{2}=r^{-2},$ and the
equation for $F$ becomes%
\begin{equation}
\frac{\partial^{2}F}{\partial r^{2}}+\frac{1}{r}\frac{\partial F}{\partial
r}-\frac{F}{r^{2}}-\frac{\partial V}{\partial F}=0
\end{equation}
Putting%
\begin{align}
F  &  =\frac{f}{a}\nonumber\\
r  &  =\frac{\rho}{a}%
\end{align}
we have%
\begin{equation}
f^{\prime\prime}+\frac{f^{\prime}}{\rho}-\frac{f}{\rho^{2}}-\frac{\partial
V}{\partial f}=0
\end{equation}
where for the HH potential $V$ we have%
\begin{equation}
\frac{\partial V}{\partial f}=-fM\left(  -1+b/2,2,16\pi^{2}f^{2}\right)
\end{equation}
where $M$ is the Kummer function. The boundary conditions are%
\begin{align}
f\left(  0\right)   &  =0\nonumber\\
f\left(  \infty\right)   &  =\text{Nonzero constant}%
\end{align}
We take \bigskip$b=1.5$, and find the numerical solution by "shooting", i.e.,
adjusting the initial conditions so as to get a nonzero $f\left(
\infty\right)  $. We obtain the desired behavior with $f(0)=0.001$,
$f^{\prime}(0)=0.2559.$ The field modulus $f\left(  \rho\right)  $ is plotted
in Fig.4. The high-energy cutoff $\Lambda$ suppresses the field at small
distances$,$ with the functional form of the field dependent on the cutoff
function. We simply set%
\begin{equation}
F\left(  r\right)  =\left\{
\begin{array}
[c]{cc}%
F\left(  \infty\right)  & \left(  r>R_{0}\right) \\
0 & \left(  r\,<R_{0}\right)
\end{array}
\right.
\end{equation}
where $R_{0}\sim\Lambda^{-1.}$ is the core radius. With this approximation,
the scalar field is uniform in a multiply-connected space.

\section{Vortex dynamics}

A simple vortex structure is the vortex ring, whose vortex line is a directed
circle of radius $R$, as illustrated in Fig.5. The ring moves normal to its
own plane, in a direction in accordance with the right-hand rule, with
velocity [25]%
\begin{equation}
v=\frac{1}{4\pi R}\ln\frac{R}{R_{0}} \label{ring}%
\end{equation}
where $R_{0}$ is proportional to the core radius. The logarithmic factor
$\ln\left(  R/R_{0}\right)  $ is slowing-varying, and may be regarded as a
constant for all practical purposes.\ Thus $v\varpropto$ $R^{-1}$
approximately. We can qualitatively understand the motion of an arbitrary
vortex line as follows. At any point on the vortex line there is a radius of
curvature $R$, which we can associate with an imaginary vortex ring of the
same radius, tangent to the line at that point. The local translational
velocity would be $v\varpropto$ $R^{-1}$ normal to this ring. The more sharply
a vortex line bends, the faster it moves perpendicular to the bending. In this
manner, a vortex line generally executes complicated self-induced motion, as
illustrated in Fig.5. The local velocity $v(s)$ of the vortex line, where $s$
is a parameter along the vortex line, is also the velocity of the superfluid
at that point.

The reconnection of vortex lines proposed by Feynman [6] is illustrated in
fig.6. It has been simulated via the nonlinear Schr\"{o}dinger equation [27].
This mechanism is important for the formation of the vortex tangle, in the
following scenario according to Schwarz [11,12]. Vortex rings will grow when
there is a normal fluid head wind, i.e., counter heat flow opposed to the
ring's translational motion, and shrink in a tail wind. Given a distribution
of vortex rings, some will grow to large sizes, and inevitably reconnect, as
schematically illustrated in Fig.7. The reconnection produces a set of smaller
rings, some of which will again grow and reconnect, and so forth, until there
is vortex tangle, like the one shown in Fig.7 through computer simulation,
with a fractal dimension 1.6 [28]. The steady-state of a vortex tangle is
maintained by a constant rate of growth and reconnections. If the heat source
is removed, the vortex tangle will decay into a sparse collection of
contracting vortex rings, and eventually disappear into the sea of quantum
fluctuations [29].

Reconnection occurs between two antiparallel vortex lines. Computer simulation
shows that parallel vortex lines tend to reorient themselves at close approach
in order to reconnect. The critical distance for reconnection between two
vortex lines with the same radius of curvature $R$ is given by [12]
\begin{equation}
\delta\approx2R\ln\frac{R}{c_{0}R_{0}} \label{del}%
\end{equation}
where $c_{0}$ is a constant. Here, the logarithmic factor is practically a
constant. Comparison with (\ref{ring}) shows $\delta\propto v^{-1}$, where $v$
is the relative velocity of the vortex segments.

As illustrated in Fig.7, reconnection creates two cusps on the newly
constituted vortex lines, with very small radii of curvature. Consequently,
the cusps will spring away from each other at very high speed, creating two
oppositely directed jets of energy, which are signature events of vortex reconnection.

In Vinen's equation $\dot{\ell}=A\ell^{3/2}-B\ell^{2}$, the coefficients $A$
and $B$ should embody all the effects discuss above. In liquid helium, $A$ is
proportional to the speed of the normal fluid. This equation has also been
derived from vortex dynamics, and $A$ and $B$ can be expressed in terms of
properties of the system of vortex lines [12]. However, they do not always
agree with the phenomenological view.

In superfluid helium, experiments reveal that the velocity distribution in the
tangle deviates from that in classical turbulence, in that it has a fat
non-Gaussian tail [24]. Reconnection events have been observed and studied
statistically [30].

%\newpage

%\section*{References}

\end{document}